\documentclass[a4paper,11pt]{article}
\pdfoutput=1
\usepackage{graphicx}
\usepackage{xcolor}
\usepackage{amsmath,amsfonts,amssymb,amstext,graphicx}
\usepackage{cancel}
\usepackage{empheq}
\usepackage{float}
\usepackage{subcaption}
\usepackage{authblk}
\usepackage{blindtext}
\usepackage{epsfig}%
\usepackage[bottom]{footmisc}
\begin{document}
\date{}

\title{Dynamical analysis of $k$-essence cosmology in the light of
Supernova Ia observations}

\author[1]{Anirban Chatterjee \thanks{Corresponding Author: anirbanc@iitk.ac.in}}
\author[2]{Abhijit Bandyopadhyay\thanks{abhi.vu@gmail.com}}
\author[2]{Biswajit Jana\thanks{vijnanachaitanya2020@gmail.com}}
\affil[1]{Indian Institute of Technology, Kanpur, Kanpur 208016, India}
\affil[2]{Ramakrishna Mission Vivekananda Educational and Research Institute, 
Belur Math, Howrah 711202, India}
\date{\today}
\maketitle

\begin{abstract}
In this paper, we analyse the JLA data on Supernova observations in the context
of $k-$essence dark energy model with Lagrangian $L=VF(X)$, with a constant
potential $V$ and the dynamical term $X = (1/2)\nabla_{\mu}\phi\nabla_{\nu}\phi = \dot{\phi}^2/2$ for a homogeneous scalar field $\phi(t)$, in a flat FRW spacetime background. Scaling relations are used to extract temporal behaviour of different
cosmological quantities and the form of the function $F(X)$ from the data.
We explore how the  parameters of the model, \textit{viz.} value of the constant potential $V$ and a constant $C$ appearing in the emergent scaling relation, control the
dynamics of the model in the context of JLA data, by setting up and analysing an equivalent dynamical system described by a set of autonomous equations.
\end{abstract}
 
\section{Introduction}
\label{sec:intro}
From the observation of type Ia Supernovae (SNe Ia),
it was first reported  independently in 1998  by  
Riess \textit{et.al.} \cite{ref:Riess98}
and Perlmutter \textit{et.al.}  \cite{ref:Perlmutter} that
 the present universe is undergoing
an accelerated expansion  and  a  
transition happened from decelerated to this accelerated phase of expansion
during late time phase of cosmic evolution of the universe.
The luminosity distances and redshifts of SNe Ia are the key observational
ingredients in establishing the features of the late-time cosmic evolution.
The present day data of   observed SNe Ia events
using diverse probes in different supernova surveys
include  various compilations  corresponding to
different redshift regions.
The small redshift $(z > 0.1)$ projects comprise  Harvard-Smithsonian Center for 
Astrophysics survey \cite{ref:Hicken}, the Carnegie Supernova Project  
 \cite{ref:Contreras,ref:Folatelli,ref:Stritzinger}  the Lick 
Observatory Supernova Search    \cite{ref:Ganeshalingam}  and the Nearby 
Supernova Factory  \cite{ref:Aldering}. SDSS-II supernova surveys 
 \cite{ref:Frieman,ref:Kessler,ref:Sollerman,ref:Lampeitl,ref:Campbell}  
are mainly focused on the redshift region of ($0.05<z<0.4$). 
Programmes like Supernova Legacy Survey  
 \cite{ref:Astier,ref:Sullivan}  the ESSENCE project \cite{ref:Wood-Vasey}, 
the Pan-STARRS survey  \cite{ref:Tonry,ref:Scolnic}  correspond to
the high redshift regime. Around one thousand SNe Ia events have been 
discovered through all surveys. In the range between $z \sim 0.01$ and $z \sim 
0.7$, luminosity distance has shown a very high statistical precision.
`Joint Light-curve Analysis  (JLA) data'  \cite{ref:Scolnic,ref:Conley,ref:Suzuki}  has been newly released, which contains total of 740 SNe Ia 
events. 
%
This entire data sample with the observed values of luminosity distances
and redshifts of SNe Ia events has been  analysed in the 
context of various cosmological studies to obtain 
features of the late-time cosmic
acceleration.\\

Dark energy, a  general label for the source of this late-time cosmic 
acceleration,  has been hypothesized as an unclustered form of energy with negative pressure - the negative pressure leading to the cosmic acceleration by counteracting the gravitational collapse.  
The phenomenological $\Lambda$-CDM model \cite{Weinberg:1988cp} of dark energy,
though fits well with cosmological data, is  plagued with
the fine tuning problem from viewpoint of particle physics.
Alternative approaches aiming construction of models of dark energy
include the  
field theoretic models \textit{viz.} 
quintessence and $k$-essence models,
in which the cosmic acceleration is driven respectively 
by scalar
fields with slowly varying potentials and  kinetic energy associated
with the scalar field through the energy-momentum 
tensor of Einstein Field equations. There are also
other viable models of dark energy based 
on modification of geometric part of Einstein’s equation ($f(R)$
gravity models \cite{fr1}), scalar tensor theories \cite{st1}, brane world models
\textit{etc.} \cite{brm1} \\

In this paper, we consider the interesting phenomenological consequences
of the $k$-essence model extracted from the SNe Ia data and 
attempted to link the extracted dynamical features of the model
with a dynamical system whose evolution is governed by a set of 
autonomous equations. The approach of using dynamical systems
in the study of cosmology has been discussed in detail in
\cite{ref:Bahamonde:2017ize,Yang:2010vv,Chakraborty:2019swx,Tamanini:2014mpa,Chatterjee:2021ijw,Dutta:2016bbs,
Barros:2019rdv,Pan:2020mst,Kase:2020hst,Amendola:2020ldb}. We assume an isotropic and homogeneous 
spacetime geometry of the universe  described by a
flat Friedmann-Robertson-Walker (FRW) metric involving
the time-dependent  scale factor $a(t)$.
The content of the universe during its late-time evolution
is approximated to be composed of dark matter and dark energy
which is consistent with the observations from Planck collaborations  
 that these two components comprise around 96\% of the
present day universe \cite{Planck:2018vyg}. The dark matter and dark energy
are modeled as mutually non-interacting ideal perfect
fluids characterised by their respective energy densities and
pressures symbolised as $(\rho_{\rm dm}, p_{\rm dm})$
and $(\rho_{\rm de}, p_{\rm de})$, with dark matter
as non relativistic dust implying $p_{\rm dm}=0$.
We consider dark energy to be represented by a homogeneous
scalar field $\phi(t)$ driven by $k$-essence Lagrangian
of the form $L = V(\phi)F(X)$, where
$X \equiv \frac{1}{2}g_{\mu\nu}\nabla^\mu\phi\nabla^\nu\phi = \frac{1}{2}
\dot{\phi}^2$ and the potential $V(\phi) = V$ is taken to be a constant.
The constancy of the $k$-essence potential ensures existence of a 
scaling relation $XF_X^2 = Ca^{-6}$ ($F_X \equiv dF/dX$ and $C$ a constant), 
which connects the scalar field $\phi(t)$ with scale factor $a(t)$.
From the model-independent analysis of the JLA SNe Ia data
we obtain the temporal behaviour of the FRW scale factor $a(t)$.
We use this in the context of our model to obtain the temporal 
behaviour of the scalar field $\phi(t)$ and consequently
extract the $X$-dependence of the dynamical term $F(X)$
in the $k$-essence Lagrangian. \\

We then show that, cosmological evolution in the context of such a 
$k$-essence model of dark energy 
(having constant potential) with the form of $F(X)$ and temporal 
behaviour of relevant cosmological quantities,
as extracted from the analysis of JLA data, can be mapped to 
the evolution of a dynamical system with properly chosen
dimensionless variables $x$ and $y$ in terms of relevant cosmological
quantities ($\phi(t), a(t)$ and its derivatives) and parameters (value of the constant
potential $V$, the constant $C$ in the scaling relation) 
of the model.  We investigate the behaviour of the
dynamical system and analyse its features
for different chosen values of involved set of parameters. This provides an indirect approach
for realising the effect of the numerical values of the constants $V$ and $C$  
in the $k$-essence cosmological model of dark energy with constant potential in the context of JLA data.\\

The paper is organised as follows.
In Sec.\ \ref{sec:data} we discussed the methodology of analysis 
of JLA data for obtaining
temporal behaviour of different relevant cosmological
quantities during the late-time phase of cosmic evolution.
In  Sec.\ \ref{sec:k} we briefly discussed the 
$k$-essence model with constant potential and used
the scaling relation to establish the connection
between the cosmological quantities and the quantities
$X$ and $F(X)$ which governs the dynamics of the $k$-essence
Lagrangian. We also presented how we used the temporal dependences of
the cosmological quantities as extracted from
the analysis of JLA data to reconstruct the form 
of the function $F(X)$. In Sec.\ \ref{sec:dyn}
we discussed the mapping of dynamical aspects of the $k$-essence
model considered along with observational inputs from
JLA data  to a two-dimensional dynamical system  driven by a set of autonomous 
equations involving  the   model parameters  $V$ and   $C$.
  The study of
fixed points of the system, based on linear stability
theory, has been presented in this section and
the implications of the values of the parameters $(C,V)$
in the determination of the fixed points have been investigated.
We summarize the conclusions of the paper 
in Sec.\ \ref{sec:con}.

\section{Cosmological parameters from JLA data}
\label{sec:data}
The recently released `Joint Light-curve Analysis'
(JLA) compilation of SNe Ia data  \cite{ref:Scolnic,ref:Conley,ref:Suzuki}, 
as discussed in Sec.\ \ref{sec:intro}
consists of luminosity distance and redshift measurements of 
740 SNe Ia events. This data set involves a compilation of SNe Ia light curves including SNe Ia data from the three-year SDSS survey, first three seasons of
the five-year SNLS survey and 14 data points in the 
very high redshift $0.7<z<1.4$ domain from HST \cite{ref:Riess}.
To take care of the different systematic uncertainties involved in the data
we analyse  the compilation of the data with flux-averaging 
technique described in \cite{ref:wang1,ref:wang2,ref:wang3}.
 The $\chi^2$-function corresponding to JLA data is given by
\begin{eqnarray}
\chi^2 = \sum_{i,j=1}^{740}(\mu_{\rm obs}^{(i)}  - \mu_{\rm th}^{(i)}) 
(\sigma^{-1})_{ij}
(\mu_{\rm obs}^{(j)}  - \mu_{\rm th}^{(j)} ) \,,
\label{eq:jla1}
\end{eqnarray}
where $\mu_{\rm th}^{(i)}$ denotes the theoretical expression for
distance modulus in a flat FRW spacetime background  at red-shift $z_i$ which is 
related to
the corresponding luminosity distance $d_L$ through 
\begin{eqnarray}
\mu_{\rm th}^{(i)} &=& 
5\log_{10} [d_L(z_{\rm hel},z_{\rm CMB})/{\rm Mpc})] + 25
\label{eq:rnew1}
\end{eqnarray}
where
\begin{eqnarray}
d_L(z_{\rm hel},z_{\rm CMB}) &=& 
(1 + z_{\rm hel}) r(z_{CMB}) \quad \mbox{with} \quad r(z) = cH_0^{-1} \int_0^z \frac{dz^\prime}{E(z^\prime)}
\label{eq:rnew2}
\end{eqnarray}
$z_{\rm CMB}$ and $z_{\rm hel}$ are SNe IA redshifts in CMB rest frame and in heliocentric frame
respectively and $H_0$ is the value of Hubble parameter at present epoch.
The observed value of distance modulus $\mu_{\rm obs}^{(i)}$ at  redshift $z_i$ 
is expressed as
\begin{eqnarray}
 \mu_{\rm obs}^{(i)} &=& m_B^\star(z_i) - M_B + \alpha X_1(z_i) - \beta C(z_i)
\label{eq:rnew3}
\end{eqnarray}
in terms of the observed peak magnitude 
$m_B^\star$, the time stretching parameter of the light-curve 
$X_1$ and supernova color at maximum brightness, $C$.
$\alpha$, $\beta$ are the nuisance parameters and $M_B$ is
the absolute magnitude kept fixed at $M_B=-19$ for the analysis performed in \cite{ ref:wang2, Bandyopadhyay:2019ukl}.
The technical details of different terms involved in $\chi^2$ and their handling
in the analysis of the data
have been comprehensively discussed in \cite{ref:wang2, ref:wang3,Bandyopadhyay:2019ukl,ref:Betoule}.
$\sigma_{ij}$ is the covariant matrix  as given in Eq.\ (2.16) of \cite{ref:wang2}. 
 The systematic uncertainties involved in the covariant matrix,
instead of dealing individually, may be handled by 
a flux averaging technique proposed by Wang in \cite{ref:wang2},
which reduces the effect of systematic uncertainties owing
to weak lensing of SNe Ia data. By this technique, one recovers
unlensed brightness of  SNe Ia events at some redshift resulting
from averaging of flux of all SNe Ia events corresponding to that redshift \cite{Wang:2003gz}.
It has also been shown in \cite{Wang:2009sn,Wang:2011sb, Wang:2005yaa} 
that the bias in distance estimation of SNe
Ia events due to systematic effects can be controlled and reduced by
this technique. 
The flux averaging technique has been comprehensively discussed in
\cite{ref:wang3} and it involves introduction of a red-shift cut-off
$z_{\rm cut}$ to separate out SN samples with
$z<z_{\rm cut}$ and $z \geqslant z_{\rm cut}$. For samples with
$z<z_{\rm cut}$, Eq.\ (\ref{eq:jla1}) has been
used to compute the $\chi^2$ and for samples with
redshifts above $z_{\rm cut}$, 
averaging of distance modulus $\mu$ and the covariant 
matrix over all the fluxes of SNe Ia samples
has been performed, and the resulting average values have
been used to compute the $\chi^2$.\\

 As mentioned in Sec.\ \ref{sec:intro}, the fact that 
the phenomenological
$\Lambda$-CDM model is plagued with
the fine tuning problem of particle physics motivates
investigation of alternative models of dark energy. A key feature of a certain class of 
such models, called  varying dark energy models, is the 
 time varying equation of
state (EOS) $w = p/\rho$ of dark energy ($\rho$ is the energy density and $p$ the pressure of dark energy). This time variation is  usually expressed in terms of variation of $w$ with redshift $z$.
The $z$- dependence of the EOS parameter $w(z)$, for the varying dark energy models,
may be constrained from the observational data. The approach involves
consideration  of various functional forms of $w(z;w_a,w_b)$, involving parameters 
$w_a$ and $w_b$ and subsequently realising the observational constraints on $w(z)$
in terms of constraints in $w_a - w_b$ parameter space. In \cite{Bandyopadhyay:2019ukl}, we have
presented the results of comprehensive analysis
of  Joint Light-curve Analysis  data  to obtain constraints in $w_a - w_b$ parameter space
 for some benchmark models -
CPL \cite{cpl-new1}, JBP \cite{jbp-new1, jbp-new2}, BA \cite{ba-new1,ba-new2}  and Logarithmic
model \cite{log-new1} - each depicting a characteristic 
functional form  of $w(z)$.   Also in \cite{ref:wang2}, Wang \textit{et. al.} performed a  
comprehensive analysis of JLA data and presented the results of analysis of their model
and performed a comparative study of results of various models.
When we consider the evolution of  universe in a FRW spacetime background during its late time phase cosmic evolution which is primarily governed by its dark matter and dark energy contents
(which constitutes 96\% of the present day universe contents),  
the reduced Hubble parameter $E(z)$ is given by
\begin{eqnarray}
E(z) & \equiv & \frac{H(z)}{H_0} = \sqrt{\Omega_{\rm dm}(1+z)^3 + \Omega_{\rm de} X(z)}
\label{eq:rnew4}
\end{eqnarray}
where $\Omega_{\rm dm}$ and $\Omega_{\rm de} \approx 1 - \Omega_{\rm dm}$
are the fractional densities of dark matter and dark energy. The dark energy density
function $X(z)$ is related to dark energy equation of state parameter $w$ as
\begin{eqnarray}
X(z) &=& \exp\left[3\int_0^z dz^\prime \frac{1+w(z^\prime)}{1+z^\prime}\right]
\label{eq:rnew5}
\end{eqnarray}
In Tab.\ \ref{tab:1} we have presented the best fit values of the
parameters $(w_a, w_b)$ obtained from the analysis of JLA data, in \cite{Bandyopadhyay:2019ukl} for  CPL, JBP, BA models and in \cite{ref:wang2} for Wang model.\\
\begin{table}[h!]
\begin{center}
 \begin{tabular}{|c|c|c|c| }
\hline
Model & $w(w_a,w_b;z)$ &  Best-fit values &  Taken from    \\
&& of $(w_a,w_b)$   &  from reference  \\
\hline
\hline
&&&\\
CPL   & $w_a + w_b \frac{z}{1+z}$ & (-0.63, -0.93) & \cite{Bandyopadhyay:2019ukl}\\
&&&\\
JBP   & $ w_a + \frac{w_bz}{(1+z)^2}$& (-0.59, -1.16) & \cite{Bandyopadhyay:2019ukl}\\
&&&\\
BA   & $  w_a + \Big{(}\frac{w_bz(1+z)}{1+z^2}\Big{)}$ & (-0.65, -0.44) & \cite{Bandyopadhyay:2019ukl}\\
&&&\\
Wang  & $ w_a \left(\frac{1-2z}{1+z}\right) + w_b \left(\frac{z}{(1+z)^2} \right)$ 
& (-1.01, -0.986) & \cite{ref:wang2}\\
\hline
\end{tabular}
\end{center}
\caption{\label{tab:1} 
Functional forms of equation
of state $w(z)$   of dark energy
expressed in terms of two parameters $w_a$ and $w_b$
as used in different  varying dark energy Models. Best fit values of the
parameters $(w_a, w_b)$ obtained from the analysis of JLA data in \cite{Bandyopadhyay:2019ukl} for (CPL, JBP, BA) and in \cite{ref:wang2} for Wang model have also been
also been presented.}
\end{table}

\begin{figure}[h]
\centering
\includegraphics[scale=0.31]{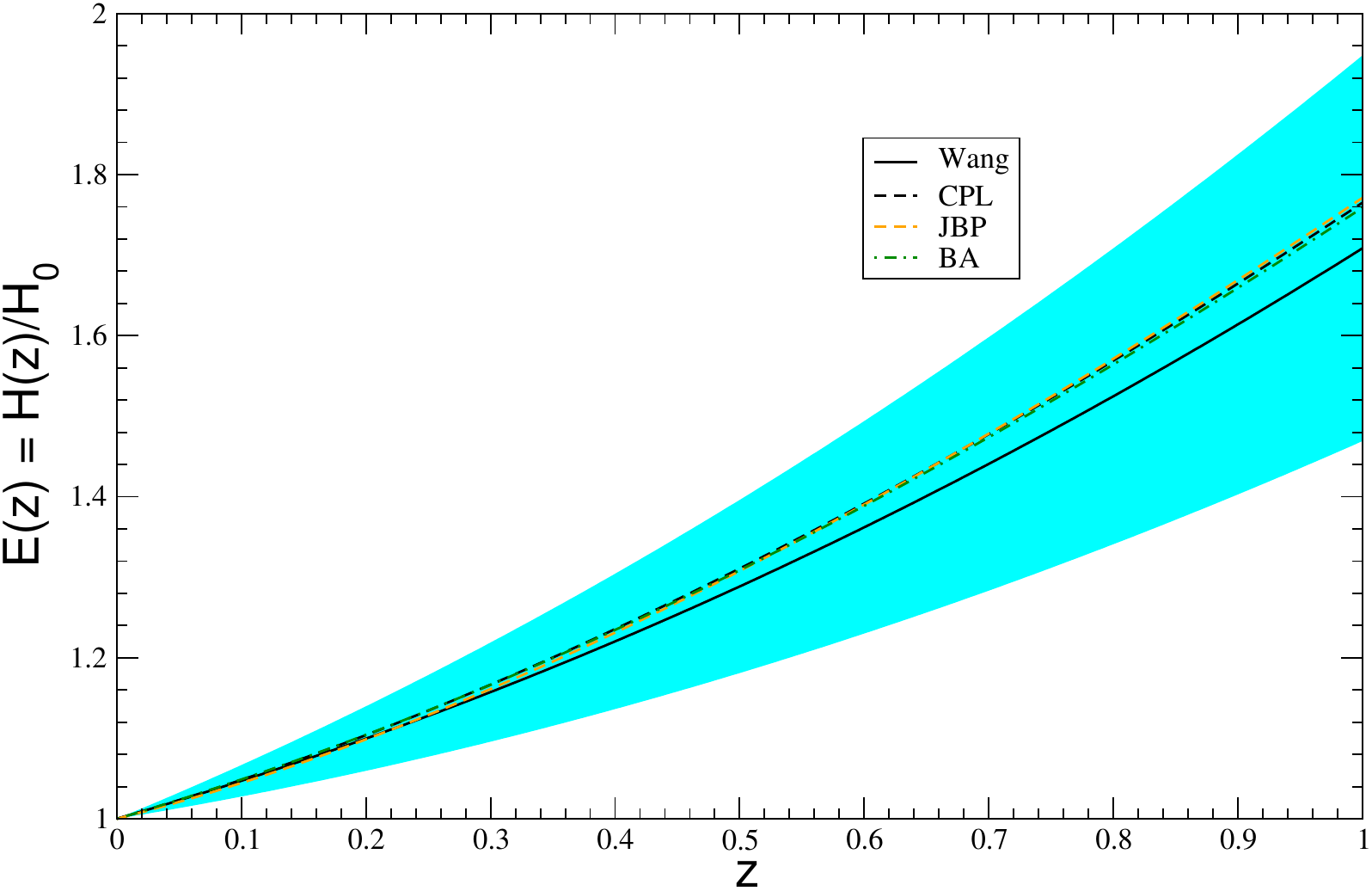} 
\includegraphics[scale=0.3]{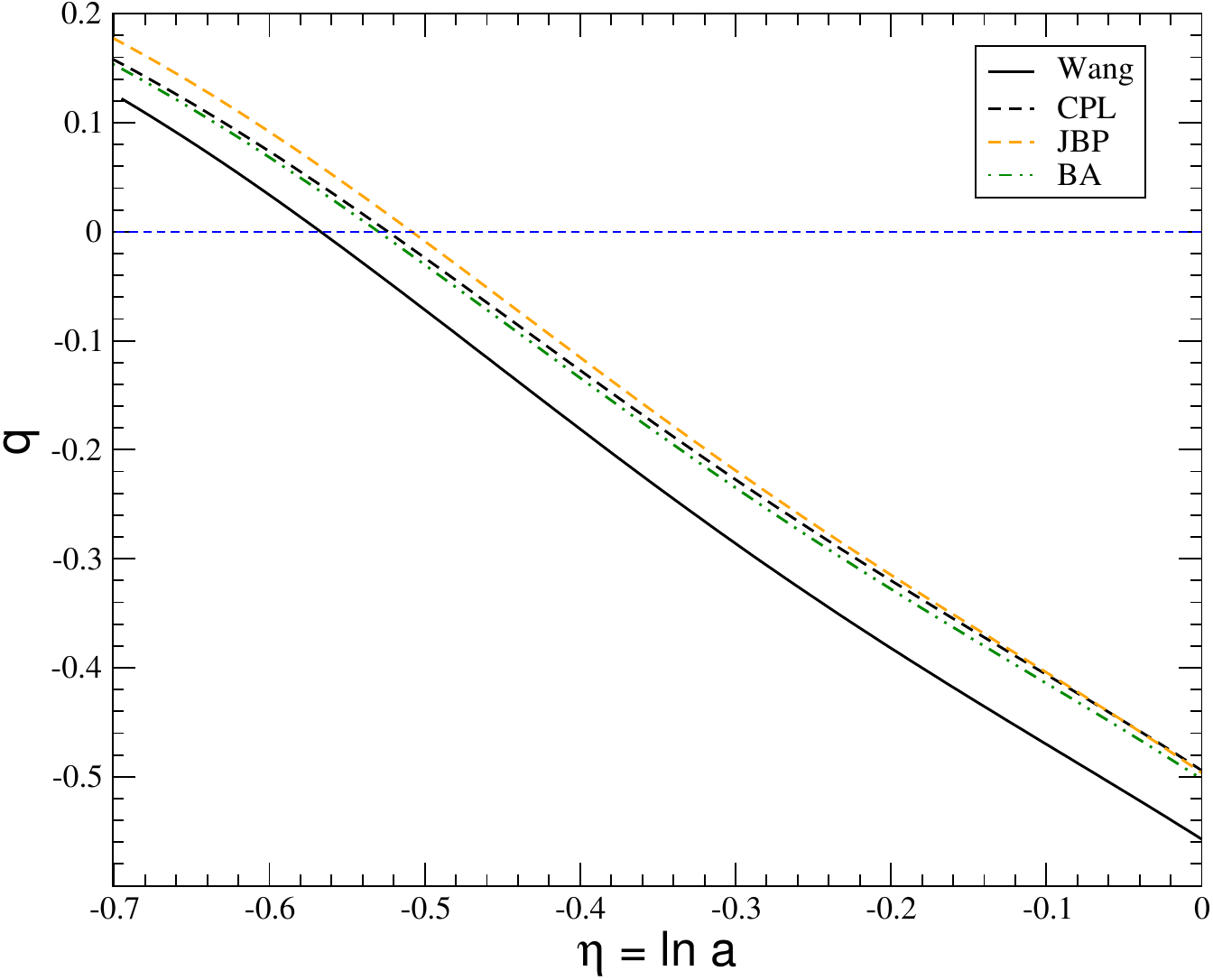} \\
\caption{ \textbf{Left panel:} 
The shaded region is the plot $1\sigma$ range of $E(z)=H(z)/H_0$ against redshift
as obtained from analysis of JLA data using Wang model presented in \cite{ref:wang2}. The solid line in middle
is the benchmark line corresponding to average of the $E(z)$-values corresponding 
to the upper  
(+$1\sigma$) and   lower   (-$1\sigma$) curves. The other dashed lines corresponds
to the corresponding curves at the best fit values of $(w_a, w_b)$ for the other 
varying dark energy models
(CPL. JBP, BA) mentioned in the text for a comparative study of different models.
In this paper, we take the $E(z)$ vs $z$ curve corresponding to
Wang model as the benchmark for
extracting temporal behaviour of other relevant cosmological quantities. \textbf{Right panel:} Plot of 
deceleration parameter $q$ as a function of  $\eta = \ln a$ at the best fit
values of $(w_a, w_b)$ corresponding to different models of variation of $w(z)$}
\label{fig:1}
\end{figure}

 The profile of the equation of state parameter $w(z)$ is translated to the profile
of the reduced Hubble constant $H(z)$ by virtue of Eqs.\ (\ref{eq:rnew4}) and (\ref{eq:rnew5}). For our analysis,
in this work, we consider the Wang model and  take the $z-$dependence of the function 
$E(z) = H(z)/H_0$ obtained in \cite{ref:wang2} from $\chi^2$-marginalisation
with respect to $M_B$ and the nuisance parameters, taking
flux averaged values of
distance modulus and the covariant matrix corresponding to 
a zero red-shift cut-off where
$H(z)$ is the Hubble parameter $\dot{a}/a$
expressed  terms of redshift and $H_0$ being its present epoch ($z=0)$ value.
 We consider the 1$\sigma$ range of the quantity $E(z)$
at every $z$,
resulting from the above analysis is shown in Fig.\ \ref{fig:1}. We obtain
the average of the $E(z)$ values in this 1$\sigma$ range for each $z$,
which is depicted by the dashed line in Fig.\ \ref{fig:1}.
We consider this central $E(z)$ vs $z$ curve  as benchmark for
extracting temporal behaviour of other relevant cosmological quantities.
 We have also obtained the 1$\sigma$ uncertainties of the cosmological
quantities corresponding to the $1\sigma$ uncertainties of $E(z)$ as shown in 
left panel of Fig.\ \ref{fig:1}.  
In the same plot, we have also depicted the $E(z)$ vs $z$ profile for the other models
at corresponding best-fit points $(w_a,w_b)$ obtained in \cite{Bandyopadhyay:2019ukl}. 
 
In the right panel of Fig.\ \ref{fig:1}, for comparison, we have also 
shown the variation of the
deceleration parameter $q = - \ddot{a}a/\dot{a}^2$, which is a dimensionless 
measure of the cosmic acceleration, as a function of a  chosen
dimensionless time  parameter
$\eta$ (chosen to depict temporal behaviour of 
relevant cosmological quantities, see Eq.\ (\ref{eq:x3})), at the best fit
values of $(w_a, w_b)$ corresponding to different models of variation of $w(z)$. \\

 The $z-$dependence of $E(z)$ as extracted from the observational data
cam be exploited to find the temporal behaviour of the FRW scale factor $a(t)$. The numerical
method of obtaining this is briefly described below.
The scale factor $a$ which is normalised to $a=1$ at present epoch,
is related to the redshift by the relation
\begin{eqnarray}
\frac{1}{a} &=& 1 + z\,.
\label{eq:rnew6}
\end{eqnarray}
Using this  we may write
\begin{eqnarray}
dt &=& -\frac{dz}{(1+z)H_0 E(z)}\,,
\label{eq:rnew7}
\end{eqnarray}
which on integration gives
\begin{eqnarray}
\frac{t(z)}{t_0} &=& 1 - \frac{1}{H_0 t_0}\int_{z}^0 \frac{dz'}{(1+z')E(z')}
\label{eq:aa1}
\end{eqnarray}
where $t_0$ denotes the present epoch. 
 Using the $E(z)$ vs $z$ profile as depicted in Fig.\ \ref{fig:1}, obtained from 
the analysis of JLA data by methodology described above,
We perform the above
integration numerically   to obtain
$z$ dependence of $t(z)$. Eqs.\ (\ref{eq:rnew6}) and (\ref{eq:aa1}) together,
provides the machinery to numerically compute simultaneous
values of $a$ and $t$ at any given redshist $z$. This amounts to 
obtaining values of $a(t)$ at corresponding $t$ eliminating $z$ from  
Eqs.\ (\ref{eq:rnew6}) and (\ref{eq:aa1}) leading to extracting
temporal behaviour of the scale factor $a(t)$ from the observational data. \\

\begin{figure}[h]
\centering
\includegraphics[scale=0.55]{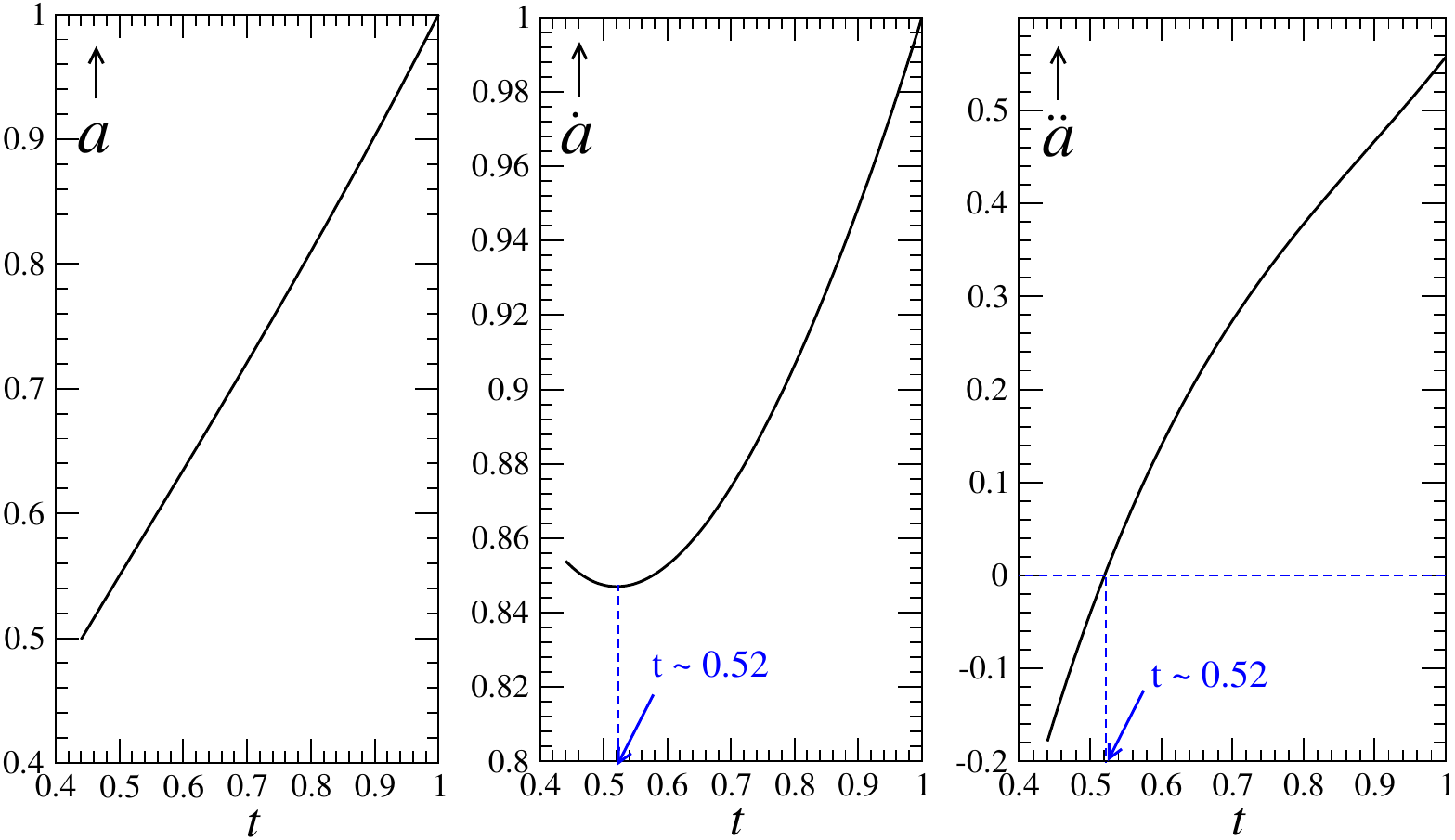} 
\caption{Plot of 
scale factor $a(t)$ (left panel), $\dot{a}(t)$ (middle panel)
and $\ddot{a}(t)$ (right panel) as a function of  $t$ corresponding
to the best-fit of Wang's model. The scale factor is normalised to
unity at the present epoch $(t=1)$. The epoch of transition from decelerated 
phase to accelerated phase of cosmic evolution ($t \sim 0.52$) is marked
on the $t-$ axis in  $\dot{a}$ vs. $t$ and  $\ddot{a}$ vs. $t$ plots in
middle panel and right panel figures respectively.} 
\label{fig:1a}
\end{figure}

 To perform this, we vary $z$  from zero (present
epoch) to $\sim 1$ (\textit{i.e.} within the accessible domain of $z$ relevant for
JLA data set)  in small steps ($\Delta z = 0.01$). We numerically
evaluate the  integral in Eq.\ (\ref{eq:aa1})
 and simultaneously compute value $a(z) \equiv
1/(1+z)$ (Eq.\ (\ref{eq:rnew6})) at each $z-$step,  
to obtain the sets of values ($t(z), a(z)$) at  each step of
values of $z$ within its above mentioned range. 
We consider scale factor $a(t)$ to be normalised to unity
at present epoch ($z=0$ or $t=1$) and found 
that, the  range $0<z<1$ corresponds 
to $t-$range: $1 > t(z) >0.44$. The obtained set of values of
 ($t(z), a(z)$) for the entire $z-$range thus gives
 variation of the scale factor with time over the time range $0.44<t<1$.
The  obtained $t-$dependence of the scale factor $a(t)$
corresponding to the best fit of the Wang model 
is shown   in left panel  of Fig.\ \ref{fig:1a}.  
  Using the obtained temporal profile of the scale factor $a(t)$, 
we used numerical differentiation to obtain the time-dependences of
the time derivatives of the scale factor, \textit{viz.}, $\dot{a}(t)$ and $\ddot{a}(t)$ and the obtained temporal profiles are respectively 
shown in middle panel and right panel of  Fig.\ \ref{fig:1a}. 
The transition from decelerated to accelerated phase of expansion
during the late time cosmic evolution, as probed by the SNe Ia observations,
is signified by the  appearance of
the minima at $t\sim 0.52$ in the time-profile of $\dot{a}$ (middle panel)
or, equivalently, by the change of sign of $\ddot{a}$ 
(transition from $\ddot{a}<0$ to $\ddot{a}>0$ regime) 
at the same epoch $(t \sim 0.52)$
in the time-profile of $\ddot{a}$ (right panel).
The temporal behaviour  of scale factor and its time derivatives
are instrumental in determining temporal profiles of
various other cosmological parameters   
like equation of state ($w$) and energy density ($\rho_{\rm dm} + \rho_{\rm de}$)
of the total dark fluid, the pressure ($P_{\rm de}$) of the dark energy fluid.
All these information, together, 
provide  the necessary observational input for 
exploring and analysing aspects of 
the $k-$essence 
model of dark energy (with a constant potential) considered in the context of
this paper.
The corresponding methodology has been comprehensively
discussed in  Sec.\ \ref{sec:k} and   \ref{sec:dyn}.\\

 We describe below, how we may exploit the observed temporal 
behaviour of temporal dependence of the scale factor to extract the 
temporal behaviour of some cosmological parameters like, equation of state
parameter $\omega$ of the total dark fluid (dark matter plus dark energy), its
total energy density  and pressure. In this context we indulge 
in a brief recollection
of the fundamental equations governing cosmological  dynamics at large scales.
The late-time cosmic evolution in a FRW spacetime background
with dark matter and dark energy as the primary content of the universe
is governed by the Friedmann equations
\begin{eqnarray}
H^2 &=& \frac{\kappa^2}{3} (\rho_{\rm dm} + \rho_{\rm de}) \label{eq:aa2}\\
\frac{\ddot{a}}{a} &=& -\frac{\kappa^2}{6} \Big{[} (\rho_{\rm dm} + \rho_{\rm de})  + 3p_{\rm de}\Big{]}\label{eq:aa3}
\end{eqnarray}
where $\kappa^2 \equiv 8\pi G$ ($G$ is the Newton's Gravitational constant), and
both dark matter and dark energy are considered
as ideal fluids characterised by their respective
energy densities and pressure: $(\rho_{\rm dm}, p_{\rm dm})$ for dark
matter and $(\rho_{\rm de}, p_{\rm de})$ for dark energy.
Besides, dark matter is considered as non-relativistic dust
implying $p_{\rm dm} = 0$. We have considered a flat spacetime  
and ignore  contributions from
radiation and baryonic matter during late time phase of cosmic evolution.
Using the above equations, 
the equation of state $w$ of the total
dark fluid (dark matter plus dark energy) can be expressed in terms of
the scale factor and its higher time derivatives as
\begin{eqnarray}
w &\equiv & \frac{p_{\rm de}}{(\rho_{\rm dm} + \rho_{\rm de} )}
= -\frac{2}{3}\frac{a\ddot{a}}{\dot{a}^2} - \frac{1}{3}
\label{eq:aa4}
\end{eqnarray}
Combining Eqs.\ (\ref{eq:aa2}) and (\ref{eq:aa3})  we obtain the continuity equation
\begin{eqnarray}
(\dot{\rho}_{\rm dm} +\dot{\rho}_{\rm de}  ) 
+ 3H (\rho_{\rm dm} + \rho_{\rm de} + p_{\rm de}) 
&=& 0
\label{eq:aa5}
\end{eqnarray}
which represents energy conservation in late time universe comprising
dark matter and dark energy. When there is no interaction between dark matter 
and dark energy, energy conservation is separately respected for both
fluids and are represented by following equations.
\begin{eqnarray}
\dot{\rho}_{\rm de} + 3H (\rho_{\rm de} + p_{\rm de}) &=& 0 \label{eq:aa6}\\
\dot{\rho}_{\rm dm} + 3H (\rho_{\rm dm}) &=& 0 \label{eq:aa7} 
\end{eqnarray}
The solution of Eq.\ (\ref{eq:aa7}) is given by 
\begin{eqnarray}
\rho_{\rm dm} &=& \rho_{\rm dm}^0 a^{-3}\,.
\label{eq:aa8}
\end{eqnarray}
We use symbols with index `0' in superscript to denote 
corresponding present-epoch values of the quantities referred by the symbols.\\

\begin{figure}[h]
\centering
\includegraphics[scale=0.45]{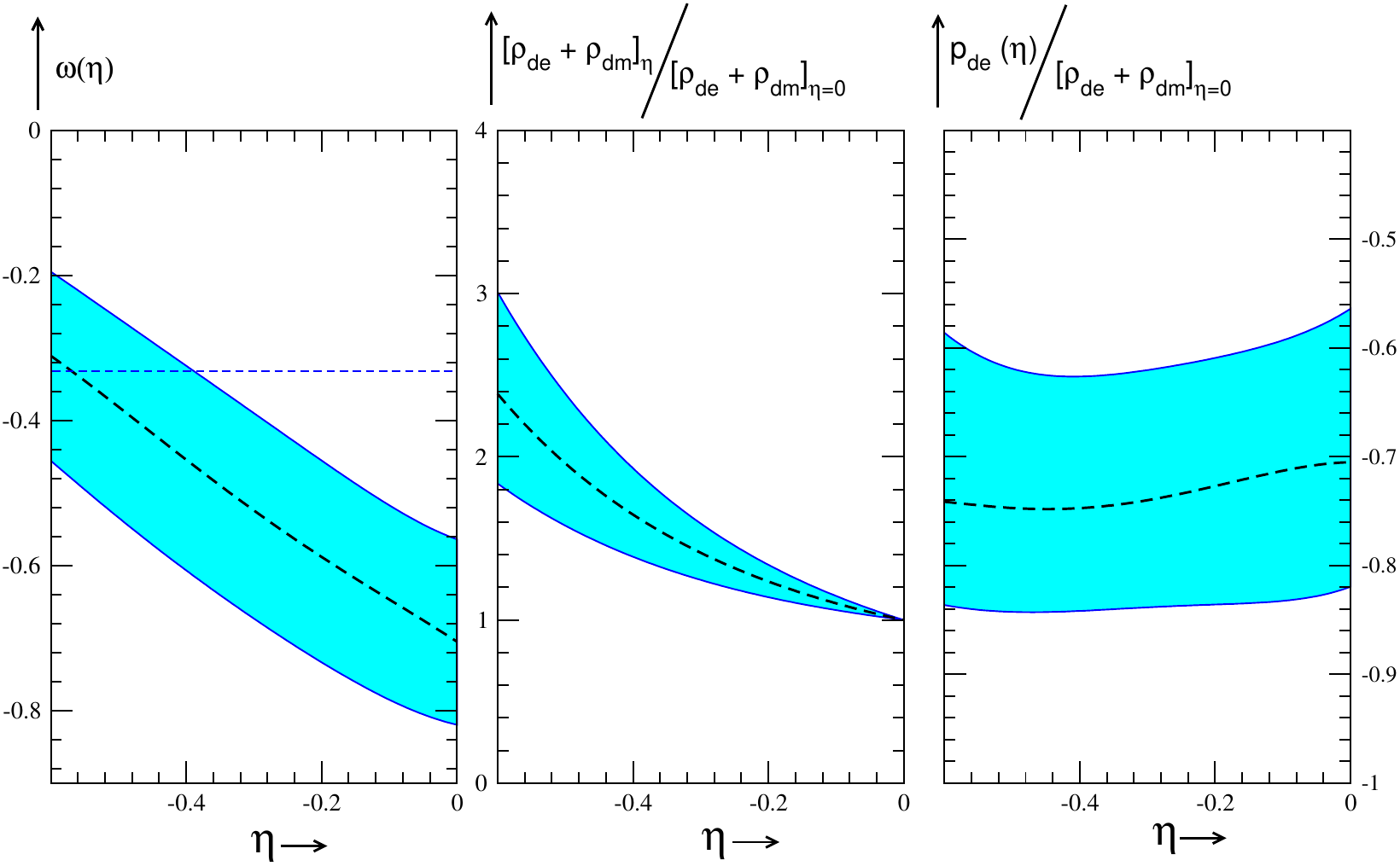} 
\caption{ Plot $1\sigma$ range of $\omega$ vs $\eta$ (shown in left panel) 
allowed from analysis of JLA data. The dashed lines
correspond to the central benchmark curve for $E(z)$  shown by dashed lines in
Fig.\ \ref{fig:1}. 
The horizontal line in the left panel represents 
the value $w = -1/3$ ($\ddot{a} =0$), corresponding to the
epoch of transition from decelerated to accelerated phase of expansion. 
Corresponding plots of 
$\frac{(\rho_{\rm de} + \rho_{\rm dm})_\eta}{(\rho_{\rm de} +
 \rho_{\rm dm})_0}$  and $\frac{(p_{\rm de})_\eta}{(\rho_{\rm de} + \rho_{\rm
  dm})_0}$ vs $\eta$ are shown in the middle panel and right panel respectively.
}
\label{fig:2}
\end{figure}

Using the temporal behaviour of the scale factor as
extracted from the JLA data, we may
exploit Eq.\ (\ref{eq:aa4}) to obtain the time dependence
of the equation of state parameter $w$ of total
dark fluid over the time domain accessible in SNe Ia
corresponding to the JLA data. The time domain as probed
in the JLA data can be expressed in terms of a
dimensionless time parameter $\eta$ as  $-0.7 <\eta < 0$ where 
$\eta$ is defined as
\begin{eqnarray}
\eta &=& \ln a \label{eq:aa9}
\end{eqnarray}
where $\eta = 0$ corresponds
to present epoch (as  $a$  at present epoch is normalised to unity).
 Note that, the   equation of state $w$ of the total dark fluid is
related to scale factor and its time derivatives by Eq.\ (\ref{eq:aa4}). 
 The obtained temporal profile of the scale factor $a(t)$ and its time derivatives
$\dot{a}(t)$ and $\ddot{a}(t)$  (presented in 
Fig.\ \ref{fig:1a})  can be used in Eq.\ (\ref{eq:aa4}) to obtain the temporal behaviour of the equation of state $w(t)$. 
Using the temporal profile of $a(t)$, we may also use Eq.\ (\ref{eq:aa9}) to 
get the relation between $\eta$ and $t$. Thus using the simultaneous values of
$w(t)$ and $\eta$ at any $t$, we can compute values of $w$ corresponding to the value of $\eta$.
The $w(\eta)$ profile, thus obtained, also expresses the temporal behaviour of
the equation of state of the total dark fluid in terms of our chosen
time parameter $\eta$.
We have chosen a suitable polynomial
to express the obtained $\eta-$dependence of $w$ by fitting the coefficients
of the polynomial with the obtained $w(\eta)$ profile.
The time-dependence of   $w(\eta)$  within its 1$\sigma$ range, 
 extracted from the analysis 
of the JLA data are shown in left panel of Fig.\ \ref{fig:2}. We find
that the temporal behaviour of $w(\eta)$, 
 corresponding to the central best-fit line in the left panel of Fig.\ \ref{fig:2},
 may be fitted with a 
  polynomial of order 5, which we express as
\begin{eqnarray}
w(\eta) &=& -1 + \sum_{i=0}B_i \eta^i \label{eq:x1}
\end{eqnarray}
with values of the coefficients $B_i$ at best-fit is given by
\begin{eqnarray}
&& B_0 = -0.70\, , B_1 = -0.61\,, B_2 = -0.49\,,
B_3 = -2.29\,,  \nonumber\\
&& B_4 = -2.81\,, B_5 = -0.92\,,  
\mbox{ and }B_i = 0\, \mbox{ for } i>5
\label{eq:x2}
\end{eqnarray}
Using the temporal behaviour of   $w(\eta)$
we also obtain the time dependence of the quantities
$(\rho_{\rm dm} + \rho_{\rm de})$ and $p_{\rm de}$ over the late time domain
$-0.7 <\eta < 0$. In terms of the parameter $\eta$ the continuity 
Eq.\ (\ref{eq:aa5}) for the total dark fluid takes the form
\begin{eqnarray}
\frac{d}{d\eta} \ln \Big{(} \rho_{\rm dm} + \rho_{\rm de}\Big{)}
&=& - 3 \Big{(} 1 + w(\eta)\Big{)}
\label{eq:x3}
\end{eqnarray}
which on integration gives
\begin{eqnarray}
\frac{(\rho_{\rm de} + \rho_{\rm dm})_\eta}{(\rho_{\rm de} + \rho_{\rm dm})_0}
&=&
\exp\left[-3\int_{0}^\eta \big{(}1 + w(\eta')\big{)}d\eta' \right] 
\label{eq:x4}
\end{eqnarray}
We use the obtained form of $w(\eta)$ as given in Eq.\ (\ref{eq:x1})
with the best-fit values of coefficients $B_i$'s (Eq.\ (\ref{eq:x2})) and perform
the integration  in the  right hand side  of Eq.\ (\ref{eq:x4}) numerically,
to obtain the total energy density as a function of $\eta$.
We find that the obtained dependence can be fitted   with an 
  order polynomial of order 5 expressed in the form 
\begin{eqnarray}
\frac{(\rho_{\rm de} + \rho_{\rm dm})_\eta}{(\rho_{\rm de} + \rho_{\rm dm})_0}
&=& \sum_{i=0}C_i\eta^i
\label{eq:x5}
\end{eqnarray}
with the best-fit values of the coefficients $C_i$'s given by
\begin{eqnarray}
&& C_0 = 1\, , C_1 = -0.89\,, C_2 = 1.28\,,
C_3 = -0.65\,,   \nonumber\\
&& C_4 = 1.36\,, C_5 = -0.97\,,  
\mbox{ and }C_i = 0\, \mbox{ for } i>5
\label{eq:x6}
\end{eqnarray}
Also from Eq.\ (\ref{eq:aa4}) we can write
\begin{eqnarray}
\frac{(p_{\rm de})_\eta}{(\rho_{\rm de} + \rho_{\rm dm})_0}
&=& w(\eta) \cdot  \frac{(\rho_{\rm de} + \rho_{\rm dm})_\eta}{(\rho_{\rm de} + \rho_{\rm dm})_0}
\label{eq:x7}
\end{eqnarray}
Using Eqs.\ (\ref{eq:x1}) and Eqs.\ (\ref{eq:x5}) we   numerically evaluated the right hand
side of the above equation for any $\eta$ and  find that, the obtained dependence fits best
with a polynomial
of order 4 expressed as
\begin{eqnarray}
 \frac{(p_{\rm de})_\eta}{(\rho_{\rm de} + \rho_{\rm dm})_0}
&=&
\sum_{i=0}D_i\eta^i
\label{eq:x8}
\end{eqnarray}
with the coefficients $D_i$'s given as
\begin{eqnarray}
&& D_0 = 0.71\, , D_1 = 0.002\,, D_2 = -0.93\,,
D_3 = -2.27\,,   \nonumber\\
&& D_4 = -1.48\,, \mbox{ and }D_i = 0\, \mbox{ for } i>4
\label{eq:x9}
\end{eqnarray}

The obtained time-dependence of $(\rho_{\rm dm} + \rho_{\rm de})$ and $p_{\rm de}$ 
thus extracted from the analysis 
of the JLA data are shown  by dashed lines
in middle and right panel of Fig.\ \ref{fig:2} by 
respectively.  The corresponding 1$\sigma$ ranges of 
the quantities have also been obtained and are shown in the 
Fig.\ \ref{fig:2} by shaded regions.

\section{$k$-essence Model with constant potential}
\label{sec:k}
We now try to realise the dynamics of dark energy 
in terms of a homogeneous scalar field $\phi(t)$ whose dynamics is driven
by a $k$-essence Lagrangian of the form $L = V(\phi)F(X)$. 
The pressure and energy density
of dark energy in this model can be expressed as
\begin{eqnarray}
p_{\rm de} &=& VF(X) \label{eq:aa10}\\
\rho_{\rm de} &=& V(2XF_X - F) \label{eq:aa11}
\end{eqnarray}
where $X = \frac{1}{2}g_{\mu\nu}\nabla^\mu\phi\nabla^\nu\phi = \frac{1}{2}\dot{\phi}^2$, $F_X \equiv dF/dX$. Using Eqs.\ (\ref{eq:aa10})
and  (\ref{eq:aa11}) in the continuity Eq.\ (\ref{eq:aa6}) of dark 
energy we obtain the equation of motion of the scalar field
as 
\begin{eqnarray}
(F_X + 2XF_{XX}) \ddot{\phi} + 3HF_X \dot{\phi} + (2XF_X - F) \frac{V_\phi}{V} 
&=& 0
\label{eq:aa12}
\end{eqnarray}
where  $F_{XX} \equiv d^2F/dX^2$, $V_\phi \equiv dV/d\phi$.
 Different variety of forms of functions $F(X)$ and $V(\phi)$
have been investigated by different authors in different contexts in 
\cite{picon-new1,picon-new2,chiba-new,scale1,scale2}. However, the simplest possible 
$k-$essence model - the purely kinetic model has been considered in \cite{scale1,scale2},
where the $k-$essence Lagrangian involves purely kinetic terms involving $X$, \textit{i.e.}
it is a function only of the derivatives of the
scalar field $\phi$ and does not depend explicitly on $\phi$. So for this class of model
$V$ in Eq.\ (\ref{eq:aa12}) is set to a constant implying $V_\phi = 0$. For this class of models
the third term of Eq.\ (\ref{eq:aa12}) vanishes and one obtains the scaling  relation \cite{scale1,scale2} 
\begin{eqnarray}
XF_X^2 &=& Ca^{-6}\, \quad , \quad C \mbox{ is a constant} 
\label{eq:aa13}
\end{eqnarray}
The existence of a scaling relation implies presence of relevant scales in the theory. This simple class of models 
were first explored as a model of inflation in \cite{picon-new3}. As elaborated in \cite{scale1}, such class of models may lead to unified dark matter with same equation of state
as that of ordinary dark matter plus a cosmological constant with an effective sound speed which is very small.
Through out the work, we have considered $V$ to be constant so that scaling is preserved.  The resulting Eq.\ (\ref{eq:aa13}) 
establishes a connection between the time derivative of the
$k-$essence scalar field ($\dot\phi$) and the
scale factor $a(t)$ whose temporal behaviour for late time 
cosmic evolution can be extracted from observed SNe Ia data. 
The scaling relation 
 is instrumental  in extracting   functional form of $F(X)$ over certain domain
 of $X$. \\

Adding  Eqs.\ (\ref{eq:aa10}), (\ref{eq:aa11}) and then
substituting $F_X$ using (\ref{eq:aa13})  we  obtain  
\begin{eqnarray}
X &=& \frac{a^6(\rho_{\rm de} + p_{\rm de})^2}{4CV^2}\,. \label{eq:aa14}
\end{eqnarray}
after some rearrangement and using Eq.\ (\ref{eq:aa10}) to Eq.\ (\ref{eq:aa14})
may be  written  as
\begin{eqnarray}
 \sqrt{X}
&=&
\left( \frac{\rho_{\rm de}^0 + \rho_{\rm dm}^0}{2\sqrt{C}V }\right)
a^3 \left[\frac{(\rho_{\rm de} + \rho_{\rm dm})}{(\rho_{\rm de}^0 + \rho_{\rm dm}^0)}
+ \frac{p_{\rm de}}{(\rho_{\rm de}^0 + \rho_{\rm dm}^0)} 
- \frac{\rho_{\rm dm}}{(\rho_{\rm de}^0 + \rho_{\rm dm}^0)}\right] \nonumber\\
&=&
\left( \frac{1}{\alpha}\right) \left[a^3 (1+w) \frac{(\rho_{\rm de} + \rho_{\rm dm})}{(\rho_{\rm de}^0 + \rho_{\rm dm}^0)} - \Omega_{\rm dm}^0\right]
=  \left( \frac{1}{\alpha}\right) g_1(\eta)
\label{eq:aa15}
\end{eqnarray} 
where we denoted $\alpha \equiv 2\sqrt{C} V / (\rho_{\rm de}^0 + \rho_{\rm dm}^0)$, 
$g_1(\eta) \equiv \left[a^3 (1+w) \frac{(\rho_{\rm de} + \rho_{\rm dm})}{(\rho_{\rm de}^0 + \rho_{\rm dm}^0)} - \Omega_{\rm dm}^0\right] $ and
$\Omega_{\rm dm}^0 \equiv
\rho_{\rm dm}^0 /  (\rho_{\rm de}^0 + \rho_{\rm dm}^0)$, the present day
fractional contribution of dark matter density to the total energy density of universe.
For numerical evaluation of various cosmological parameters we used
the observed value   $\Omega_{\rm dm}^0 = 0.268$ from  Planck observation \cite{Planck:2018vyg}. 
Again using Eq.\ (\ref{eq:aa13}) to (\ref{eq:aa15}) in  Eq.\ (\ref{eq:aa10})  
we can write
\begin{eqnarray}
  \frac{F(X)}{2\sqrt{C}}   &=&   \left(\frac{1}{\alpha}  \right)
\frac{w(\rho_{\rm de} + \rho_{\rm dm})}{ (\rho_{\rm de}^0 + \rho_{\rm dm}^0)}
= \left(\frac{1}{\alpha}  \right) g_2(\eta)
\label{eq:aa16}
\end{eqnarray}
where $g_2(\eta) \equiv \frac{w(\rho_{\rm de} + \rho_{\rm dm})}{ (\rho_{\rm de}^0 + \rho_{\rm dm}^0)}$. Note that, temporal behaviour of  all
the quantities involved in the functions $g_1(\eta)$ and $g_2(\eta)$
occurring in Eqs.\ (\ref{eq:aa15}) and (\ref{eq:aa16}), have
already been obtained from analysis of JLA data for the 
late time phase of cosmic evolution, as discussed in
Sec.\ \ref{sec:data}. From the above two equations we have
$\frac{F(X)}{\sqrt{X}} = \sqrt{2C} \frac{g_2(\eta)}{g_1(\eta)}$
which implies that chosen value of the constant $C$ sets
a constant scaling to time profile of the
quantity $\frac{F(X)}{\sqrt{X}}$. The value of $\frac{F(X)}{\sqrt{X}}$
is however independent of $\alpha$ (\textit{i.e.} $V$). However, 
by eliminating the time parameter  $\eta$ froms Eqs.\ (\ref{eq:aa15})
and (\ref{eq:aa16}), we may get the functional form of $F(X)$ as a function of $X$.
This would require knowledge of the inverse function $g_1^{-1}$ of $g_1$, since
inverting Eq.\ (\ref{eq:aa15}) we can write, 
$\eta = g_1(\alpha \sqrt{X})$ and putting it in
Eq.\ (\ref{eq:aa16}) we have
\begin{eqnarray}
\frac{F(X)}{\sqrt{2C}} 
&=& \left(\frac{1}{\alpha}\right) g_2 \left(g_1^{-1}(\alpha \sqrt{X})\right)\,.
\label{eq:rev10}
\end{eqnarray}
So, both $C$ and $\alpha$ (\textit{i.e.} both $C$ and $V$) plays the roll
of parameters in the functional form of $F(X)$. Using our knowledge of
the numerical values of the functions $g_1(\eta) (= \alpha\sqrt{X})$ 
and $g_2(\eta) \left(= \alpha \frac{F(X)}{\sqrt{2C}}\right)$
at different values of $\eta$ as extracted from analysis of JLA data
discussed above, we can obtain the numerical values     
of $F(X)/2\sqrt{C}$ for different values of $X$ over a certain
domain, for specific choices
of the values of constant $\alpha$. \\

\begin{figure}[h]
\centering
\includegraphics[scale=0.5]{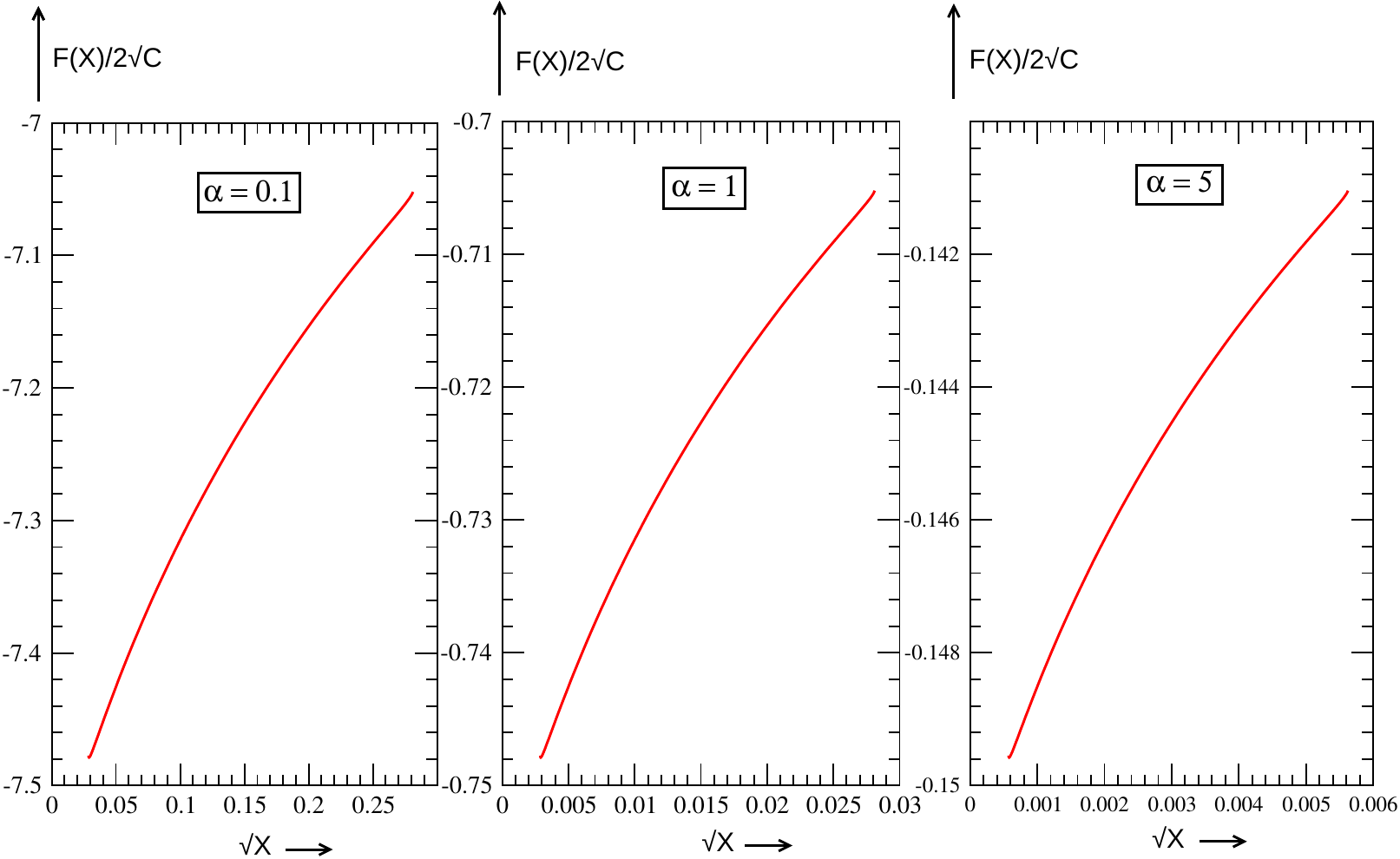} 
\caption{Profile of $F(X)/2\sqrt{C}$ 
obtained from the analysis 
of JLA data corresponding to the benchmark temporal profile
shown by the dashed  lines in plots of Fig.\ \ref{fig:2}}
\label{fig:3}
\end{figure}

To depict the dependence of $F(X)$ on $\sqrt{X}$ as extracted from the observational
data for constant potential $k$-essence scenario, we choose 
three benchmark values of the constant $\alpha$ \textit{viz.} 0.1, 1 , 5.
For each choice, we have shown the profile of   $F(X)/2\sqrt{C}$ 
obtained from the analysis 
of JLA data in Fig.\ \ref{fig:3}. The obtained dependence is found to follow the profile 
of a polynomial of
$\sqrt{X}$ of degree 3 
\begin{eqnarray}
\frac{F(X)}{2\sqrt{C}} &=& A_0(\alpha) + A_1(\alpha) \sqrt{X} + A_2(\alpha) (\sqrt{X})^2
+ A_3 (\alpha)(\sqrt{X})^3  \label{eq:aa17}
\end{eqnarray}
where the best-fit values of coefficients, determined for  
the  three benchmark cases and are 
presented  in Table\ \ref{tab:A}.\\

\begin{table}[h]
	\centering
	\begin{tabular}{|c|c|c|c|c|}
		\hline 
		$\alpha$ & $A_0(\alpha)$ & $ A_1(\alpha) $ & $A_2(\alpha)$ & $ A_3(\alpha) $  \\
		\hline
		0.1 & -7.6 & 0.4 & 2.24 & -1.97\\
		\hline
		1 & -0.76 & 0.28 & 0.88 & -2.53\\
		\hline
		5 & -0.15 & 0.056 & 2.24 & -13.97\\
		\hline
    	\end{tabular}
\caption{Values of coefficients in Eq.\ (\ref{eq:aa17}) obtained from the analysis of JLA data,
for different choices of values of the constant $\alpha$.}
\label{tab:A}
\end{table} 
With the form of $F(X)$ in Eq.\ (\ref{eq:aa17}) 
 along with the values of coefficients $A_i$'s as 
extracted from the analysis of JLA data,
we write down the pressure and energy density of dark energy from 
Eqs.\ (\ref{eq:aa10}) and\ (\ref{eq:aa11}) in terms of $X$ as
\begin{eqnarray}
\frac{p_{\rm de}}{(\rho_{\rm de}^0 + \rho_{\rm dm}^0)} 
 &=& \alpha  \left(A_0(\alpha) + A_1(\alpha) X^{\frac{1}{2}} + A_2(\alpha) X
+ A_3 (\alpha)X^{\frac{3}{2}} \right) \label{eq:aa18}\\ 
\frac{\rho_{\rm de}}{(\rho_{\rm de}^0 + \rho_{\rm dm}^0)}&=& \alpha   \left(-A_0(\alpha) + A_2 (\alpha) X + 2 A_3(\alpha) X^{\frac{3}{2}} \right) \label{eq:aa19} 
\end{eqnarray}

\section{Aspects of the $k$-essence model in terms of a dynamical system}
\label{sec:dyn}
The $k$-essence model 
of dark energy with constant potential considered here 
involves two parameters $C$  and $V$.
We explore how these parameters   control 
the cosmological dynamics of the model in the context of JLA data 
 from the perspective of dynamical system analysis.  
The technique involves a suitable choice of two  
dimensionless dynamical variables $x(\eta)$ and $y(\eta)$ 
in terms of relevant cosmological quantities and parameters of
the theory and converting equations representing cosmological
dynamics into  an autonomous system of ordinary differential
equations of the form
\begin{eqnarray}
x^\prime = F_1(x,y) &\mbox{ and }& y' = F_2(x,y)
\label{eq:auto}
\end{eqnarray}
where $^\prime$ represents the derivative with respect to the
time parameter $\eta$. $F_i(x,y)$ ($i=1,2$) only depend on $x,y$
without having any explicit time-dependence.
The fixed points $(x_0,y_0)$ of the system of Eqs.\ (\ref{eq:auto})
correspond to  
$F_i(x_0,y_0) = 0$ ($i=1,2$). Stability of the stationary points 
may be analysed by the method based on linear stability theory.
This involves Taylor expansion of $F_i(x,y)$ around the fixed point
$(x_0,y_0)$ which requires knowledge of the corresponding
Jacobian matrix $J$ whose elements $J_{ij}$ are the first order
derivatives of $F_i(x,y)$, given by
\begin{eqnarray}
J_{ij} &=& \frac{\partial F_i}{\partial x_j}\quad \mbox{ with } x_1=x\,, x_2=y
\label{eq:jaco}
\end{eqnarray}
In the context of linear stability theory,
the nature of the fixed points of the system can be 
broadly classified into three categories depending on 
nature of eigenvalues of the Jacobian matrix, $J$ evaluated at the fixed
points, provided all the eigenvalues 
have non-zero real parts. In the scenario discussed here, $J$ is a $2\times 2$  
matrix with two eigenvalues.
If real parts of all the non-zero eigenvalues
of $J$ at a fixed point are all negative, 
such a fixed point attracts all its
nearby trajectories in $x-y$ plane and is referred as
a stable fixed point or attractor. If all the non-zero eigenvalues
evaluated at a fixed point
have positive real parts, the  nearby trajectories
of the fixed point are repelled from it and the fixed
point is referred as an unstable fixed point or repeller.
If the two eigenvalues of the $J$ at any fixed point
have real parts with mutually opposite signs, its called
a saddle point and in $x-y$ plane it attracts some of its nearby
trajectories and repels others. In cases, where any (or more) of the
eigenvalues have zero real part the linear stability theory fails to
explore the nature of the fixed points.\\

To map the dynamics of $k$-essence model to a system
of autonomous equations  we define the two dimensionless
dynamical variables $x$ and $y$ as 
\begin{eqnarray}
x = \dot{\phi} \quad \mbox{and} \quad y =  \frac{\kappa \sqrt{V}}{\sqrt{3}{H}} \label{eq:defxy}
\end{eqnarray}
Using Eq.\ (\ref{eq:aa12}) (with $V_\phi=0$ for constant potential $V$), 
Eq.\ (\ref{eq:aa17}) and the fact that for a homogeneous 
field $\phi(t)$, $X = (1/2)\dot{\phi}^2$,
the temporal evolution of the quantity $x=\dot{\phi}$ with
respect to the  time parameter $\eta = \ln (a)$ may be written as
\begin{eqnarray}
x^\prime  &=& -3  \left(\frac{A_1(\alpha) + \sqrt{2}A_2(\alpha)  x+\frac32 A_3(\alpha)  x^2}{\sqrt{2}A_2(\alpha)  +3A_3(\alpha) x } \right) \label{eq:xp} 
\end{eqnarray}
Derivative of $y$ with respect to $\eta$ gives $y^\prime = -y\dot{H}/H^2$. Using  
Eqs.\ (\ref{eq:aa2}), (\ref{eq:aa3}) and  (\ref{eq:aa18}) we may write 
this time evolution of $y$ as
\begin{eqnarray}
y^{\prime} &=&  \frac{3}{2} y \left[1+\frac{y^2\sqrt{C}}{\sqrt{2}} 
\left(2\sqrt{2}A_0(\alpha) +2A_1(\alpha) x+\sqrt{2}A_2(\alpha) x^2+A_3(\alpha) x^3\right) \right]\label{eq:yp}
\end{eqnarray}
As discussed in Sec.\ \ref{sec:data},   
the analysis of data from  SNe IA observations
is instrumental in capturing the 
observable features of late time cosmic evolution of the universe
in the parameters $A_i(\alpha)$'s arising in the context
of the $k$-essence model considered here. 
Observed features of temporal behaviour of
quantities like $\rho_{de}$, $P_{\rm de}$,  $F(X)$ \textit{etc.}, 
relevant in this context,
have been expressed through their respective expressions
involving the parameters  $A_i(\alpha)$'s, 
as presented  in Sec.\ \ref{sec:k}. The dynamical
equations which have been set up based on this, therefore, contain
observational inputs as extracted from the JLA data.\\

Note that, though the $x^\prime$-equation (\ref{eq:xp})
does not contain $y$ explicitly, the two autonomous 
Eqs.\ (\ref{eq:xp}) and (\ref{eq:yp}) 
representing the dynamical system are  coupled through the 
parameter $\alpha$. The value of the parameter 
$\alpha$, 
enters in the Eqs.\ (\ref{eq:xp}) and (\ref{eq:yp}) through the 
coefficients $A_i$'s. Also, the $y^\prime$-equation (\ref{eq:yp}) 
contains the parameter $C$, apart from $\alpha$.
The parameter $\alpha \equiv 2\sqrt{C} V/(\rho_{\rm de}^0 + \rho_{\rm dm}^0)$,
in turn, is determined by the choice of
the value of the constant $C$ occurring in
the scaling relation\ (Eq.\ \ref{eq:aa13}) and also on
the constant value of the potential $V$ in the $k$-essence model
considered here. So choice of the parameters $(C,V)$ or
equivalently ($C,\alpha)$ determines the set of autonomous
Eqs.\ (\ref{eq:xp}) and (\ref{eq:yp}) describing the dynamical system.\\

\begin{table}[h]
\centering
\begin{tabular}{ |c|l|c||l|c||l|c||}
  \hline
  & \multicolumn{2}{|c||}{$\alpha=0.1$} & \multicolumn{2}{|c||}{$\alpha=1.0$}
  & \multicolumn{2}{|c||}{$\alpha=5.0$}\\
 \cline{2-7}
$C$ & Fixed Points & Stability &  Fixed Points & Stability
&  Fixed Points & Stability \\
 \hline
 \hline
& (-0.11,-0.45) & Stable & (-0.16,-1.42) & Stable & (-0.01,-3.22) & Stable\\
  &(-0.11,0) & Saddle & (-0.16,0)  & Saddle  & (-0.01,0) & Saddle \\
0.1  & (-0.11,0.45) & Stable & (-0.16,1.42)  & Stable & (-0.01,3.22) & Stable \\
  &(1.18,-0.48)  & Stable & (0.47,-1.53) & Stable & (0.17,-3.4)  & Stable   \\
  &(1.18,0)   & Saddle  & (0.47,0)   & Saddle & (0.17,0) & Saddle\\
  &(1.18,0.48)  & Stable & (0.47,1.53)  & Stable & (0.17,3.4) & Stable  \\
 \hline
 & (-0.11,-0.25) & Stable & (-0.16,-0.9) & Stable & (-0.01,-1.81)& Stable\\
  & (-0.11,0)  & Saddle &(-0.16,0) & Saddle & (-0.01,0) & Saddle  \\
1  & (-0.11,0.25) & Stable & (-0.16,0.9) & Stable & (-0.01,1.81) & Stable  \\
 & (1.18,-0.27) & Stable &  (0.47,-0.86)  & Stable & (0.17,-1.91) & Stable  \\
  & (1.18,0)  & Saddle &  (0.47,0)   & Saddle & (0.17,0)  & Saddle \\
  & (1.18,0.27)  & Stable & (0.47,0.86)  & Stable & (0.17,1.91) & Stable  \\
 \hline
& (-0.11,-0.14) & Stable & (-0.16,-0.45) & Stable & (-0.01,-1.01)& Stable \\
  & (-0.11,0) & Saddle  & (-0.16,0) & Saddle & (-0.01,0)  & Saddle\\
10 & (-0.11,0.14) & Stable & (-0.16,0.45) & Stable& (-0.01,1.01)  & Stable  \\
  & (1.18,-0.15) & Stable & (0.47,-0.48)   & Stable & (0.17,-1.07)& Stable\\
  & (1.18,0)  & Saddle & (0.47,0)  & Saddle & (0.17,0) & Saddle\\
  & (1.18,0.15)& Stable & (0.47,0.48)  & Stable &(0.17,1.07)  & Stable \\
 \hline
\end{tabular}
\caption{Fixed points and their stability for the dynamical system represented by 
set of Eqs.\ (\ref{eq:xp}) and (\ref{eq:yp}) corresponding to 
different benchmark choices of $(C,\alpha)$.}
\label{table:2}
\end{table}
\begin{figure}[h]
\centering
\includegraphics[scale=0.55]{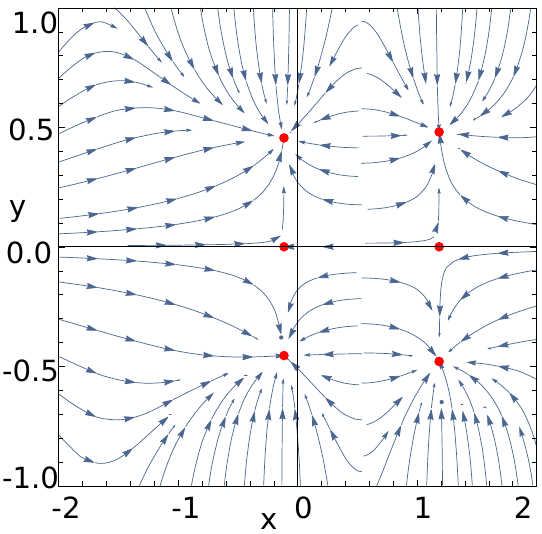} 
\includegraphics[scale=0.55]{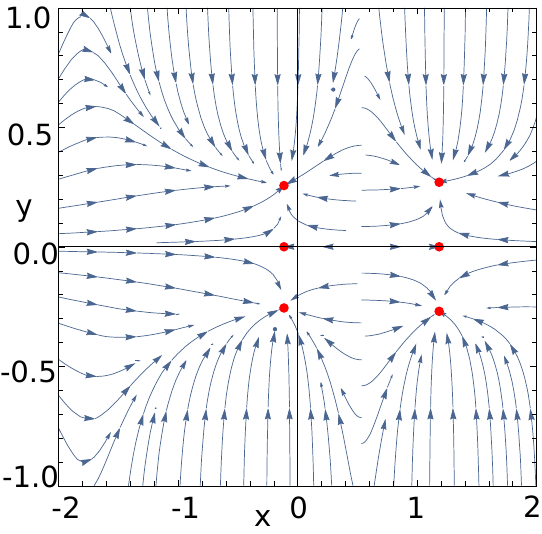} 
\includegraphics[scale=0.55]{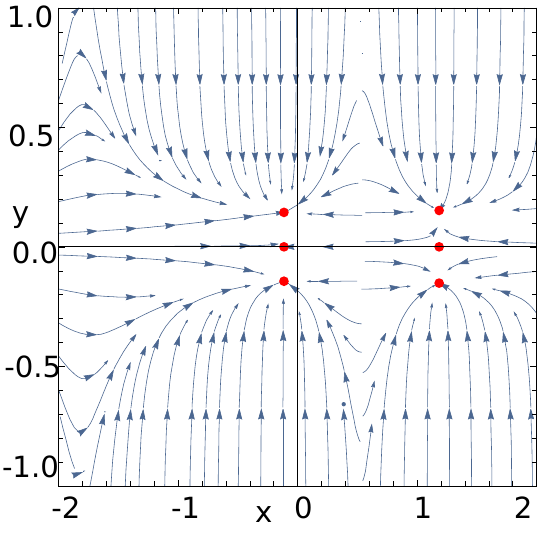} \\
\caption{Fixed points and orbits in the $x-y$ plane 
for $\alpha=0.1$ with $C=0.1$ (left panel),
$C=1$ (middle panel), $C=10$ (right panel)}
\label{fig:ppap1}
\end{figure}
\begin{figure}[h]
\centering
\includegraphics[scale=0.55]{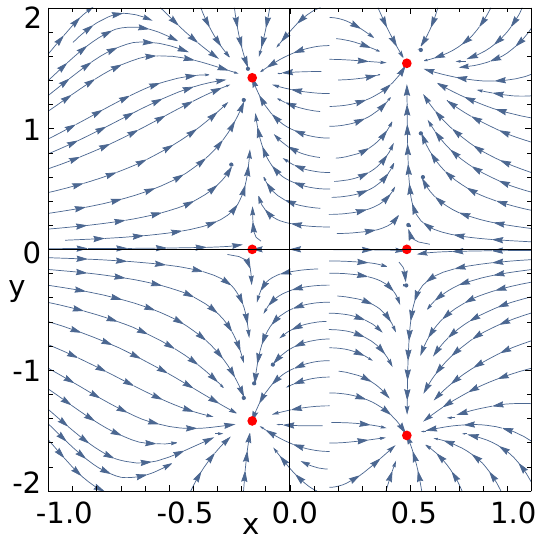} 
\includegraphics[scale=0.55]{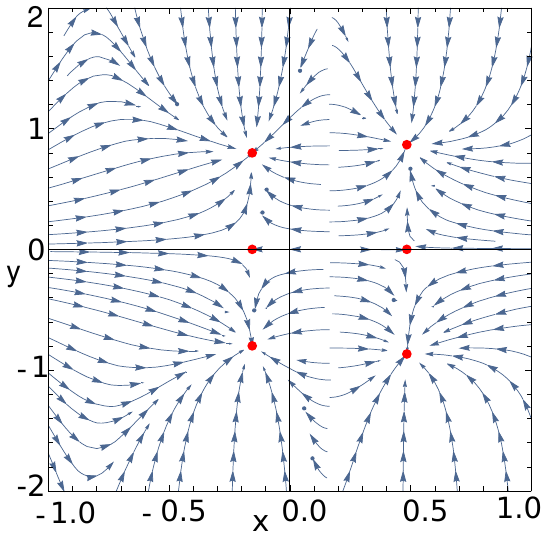} 
\includegraphics[scale=0.55]{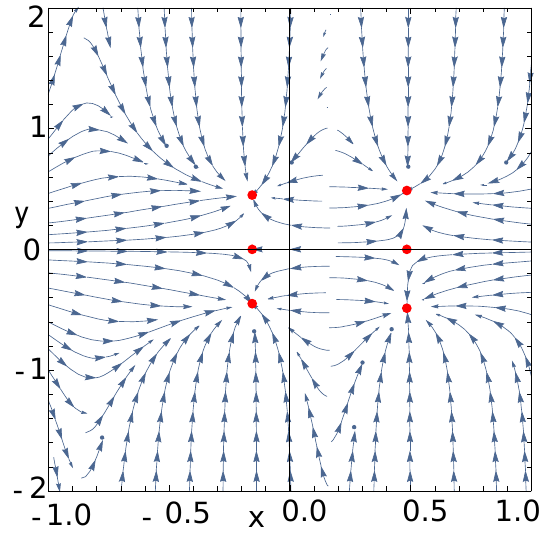} \\
\caption{Fixed points and orbits in the $x-y$ plane 
for $\alpha=1$ with $C=0.1$ (left panel),
$C=1$ (middle panel), $C=10$ (right panel)}
\label{fig:ppa1}
\end{figure}
\begin{figure}[h]
\centering
\includegraphics[scale=0.55]{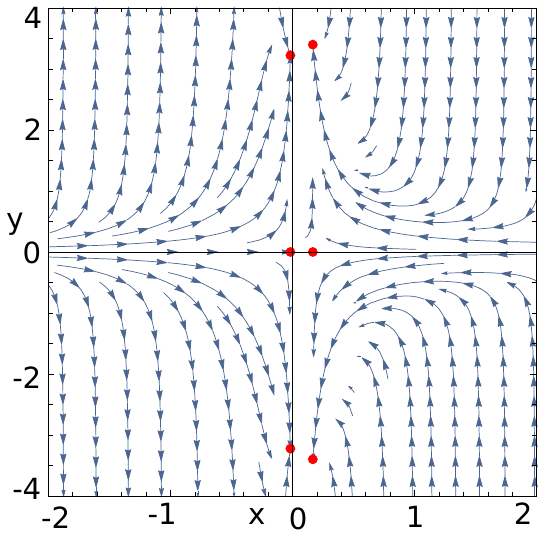} 
\includegraphics[scale=0.55]{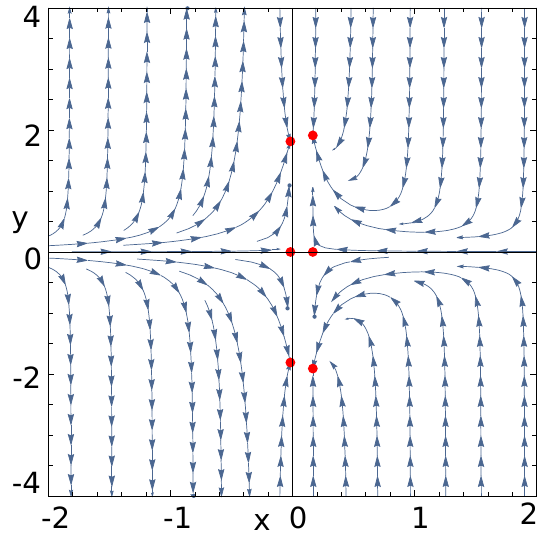} 
\includegraphics[scale=0.55]{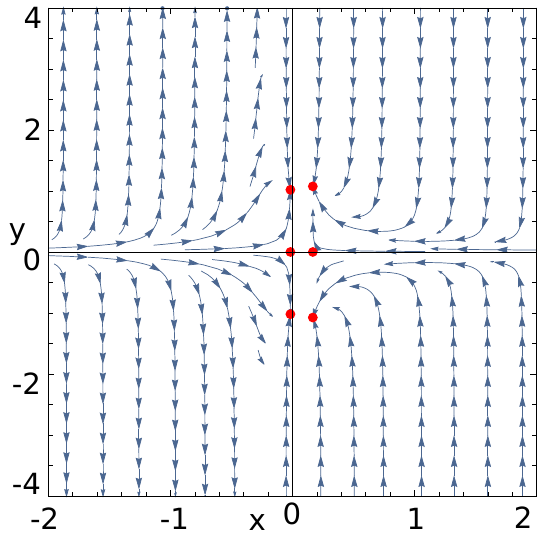} \\
\caption{Fixed points and orbits in the $x-y$ plane for $\alpha=5$ with $C=0.1$ (left panel),
$C=1$ (middle panel), $C=10$ (right panel)}
\label{fig:ppa10}
\end{figure}

The dynamical system governed by
set of autonomous equations\ (\ref{eq:xp}) and\ (\ref{eq:yp}) represents
the  motion of a hypothetical 
particle moving in two-dimension  described in terms of 
Cartesian coordinates $x(\eta)$ and $y(\eta)$. The  
dynamics of cosmic
evolution gets captured into the aspects of motion of the
hypothetical particle through defining expressions of  $x$ and $y$  in terms
of cosmological variables and parameters  (Eq.\ (\ref{eq:defxy})).
This approach of dynamical system analysis
in the context of late time cosmic evolution
with  dark energy dynamics represented by a $k$-essence scalar
field driven by a constant potential offers a way
to explore how the features of the late time cosmic evolution 
depends  on the  numerical values of the constants $C$ and  $V$,  within
the context of JLA observational data.\\

For illustration, the fixed points  and  nature of their stability 
for some benchmark 
choices of values of the parameter set ($C,\alpha$) are presented in
Table\ \ref{table:2}.  
Our analysis shows that the number of fixed points
of the system and their respective nature  are insensitive 
to the values of $C$ and $\alpha$.   
However, the coordinates of fixed points
change with the change of values of these parameters.  
For different choices of values of  ($C,\alpha$)
the system is always found to have six different fixed points out of
which four are stable fixed points and two are  saddle points.
The two saddle fixed points 
(two eigenvalues of the Jacobian having mutually opposite
signs) are always found to lie on the $x$-axis ($y$=0):
 one on the positive side and the other on the negative side.
Both the saddle points approach more closer to the origin with 
increasing values of $\alpha$ but do not move with change of $C$
for a fixed $\alpha$. This can be seen from the 
Table \ref{table:2} as well as from the Figs.\  \ref{fig:ppap1},
\ref{fig:ppa1} and \ref{fig:ppa10}. Each of 
four stable fixed points  lie on each quadrant of the $x-y$ plane.
The stable points in the 1st and 4th quadrant
are reflections of each other with respect to $x-$axis (same $y$
values with opposite signs). Same is true for the stable points
in 2nd and 3rd quadrant. Each of the stable points move towards
the $x-$axis with increasing values of $C$ for a given $\alpha$
and move towards the  $x-$axis with increasing values of $\alpha$ 
for a given $C$.\\

Analytically we see from   Eq.\ (\ref{eq:yp})
that, $y^\prime = 0$ corresponds to either of the two equations:
\begin{eqnarray}
&& y_0 = 0 \label{eq:ysol1}\\
&&  
y_0^2 \left(2\sqrt{2}A_0(\alpha) +2A_1(\alpha) x_0 + \sqrt{2}A_2(\alpha) x_0^2 + A_3(\alpha) x_0^3\right)   = - \frac{\sqrt{2}}{\sqrt{C}} \label{eq:ysol2}
\end{eqnarray}
The saddle fixed points always correspond to the $y_0=0$ solution (Eq.\ (\ref{eq:ysol1})) and lie on the $x-$axis. The stable fixed points
always correspond to the solution (Eq.\ (\ref{eq:ysol2}))
which has a $y_0 \to -y_0$ symmetry justifying occurrence of
each pair of fixed points as reflections of each other about the
$x-$axis. Again from  Eq.\ (\ref{eq:xp}), $x^\prime = 0$ 
corresponds to the quadratic equation
\begin{eqnarray}
 A_1(\alpha) + \sqrt{2}A_2(\alpha)  x_0 + \frac{3}{2} A_3(\alpha)  x_0^2 &=& 0 
 \label{eq:xsol1} 
\end{eqnarray}
whose two solutions are
\begin{eqnarray}
x_0 &=& \frac{-\sqrt{2}A_2(\alpha) \pm \sqrt{2A_2(\alpha)^2 - 6A_1(\alpha) A_3(\alpha)}}
{3 A_3(\alpha)} \label{eq:xsol2} 
\end{eqnarray}
For each of two solutions for $x_0$, the solutions of the Eq.\ (\ref{eq:ysol2})
results in four stable fixed points of the system in $x-y$ plane.
The  positions of the stable points
in the $x-y$ plane are controlled by the value of $C$
appearing in Eq.\ (\ref{eq:ysol2}) and the value of $\alpha$ 
entering both the solutions in Eq.\ (\ref{eq:ysol2}) and Eq.\ (\ref{eq:xsol2}) 
through the quantities $A_i(\alpha)$s. 
We also observe that the constant $C$ appears in the 
$y^\prime$ equation (\ref{eq:yp}) as $\sqrt{C}$. So
for any given  choice of $C>0$, the 
 negative value of its
square root has to be considered separately for
a complete analysis. 
For different above mentioned choices of values of $C$ and $\alpha$, 
consideration of the  negative  
square root ($\sqrt{C}<0$) in Eq.\ (\ref{eq:yp}),
gives  only two real fixed
points, both of which are  found to be saddle points corresponding
to $y_0=0$ fixed point solution.\\

Also note that, using Eqs.\  (\ref{eq:aa4}), (\ref{eq:aa10}), (\ref{eq:aa17}), 
 (\ref{eq:defxy}) and $X = \dot{\phi}^2/2$, 
 the equation  of state of the total dark fluid may be expressed 
as
\begin{eqnarray}
\omega &=&  
   \sqrt{C} y^2 \left[2 A_0(\alpha) +  \sqrt{2} A_1(\alpha) x +  A_2(\alpha)  x^2 
+ \frac{1}{\sqrt{2}} A_3 (\alpha) x^3 \right]  \label{eq:omegaxy}
\end{eqnarray}
Therefore,  for all the saddle fixed points ($y_0=0$)
we have $w = 0$ which does not correspond to the 
acceleration of the present universe as revealed
from Supernova Ia observations. For
all the other obtained real fixed points, we find, $w = -1$
implying all the stable fixed points or attractors correspond
to an accelerating universe.\\

 Note that using Eq.\ (\ref{eq:aa14}) we can express the scale factor $a$ in terms of the kinetic term $X = \dot{\phi}^2/2$ of
the $k-$essence model as
\begin{eqnarray}
a &=& \left[\frac{4CV^2 X}{(\rho_{\rm de} + p_{\rm de})^2}\right]^{1/6}
\label{eq:ab1}
\end{eqnarray}
Using Eqs.\ (\ref{eq:aa18}) and (\ref{eq:aa19}) in Eq.\ (\ref{eq:ab1}) and putting
$\alpha \equiv 2\sqrt{C} V / (\rho_{\rm de}^0 + \rho_{\rm dm}^0)$ we get,
\begin{eqnarray}
a &=& \left[\frac{\sqrt{X}}{A_1(\alpha) \sqrt{X} + 2A_2(\alpha) X + 3A_3(\alpha) X^{3/2}}\right]^{1/3}
\label{eq:ab2}
\end{eqnarray}
Since $x = \dot{\phi} = \sqrt{2X}$, the above equation establishes connection between
the dynamical variable $x$ and scale factor $a$ for any chosen values of $\alpha$
(\textit{i.e.} the constants $C$ and $V$). The other dynamical variable, $y$, 
as defined by Eq.\ (\ref{eq:defxy}), is related to Hubble Parameter
$H \equiv \dot{a}/a = (d/dt) \ln a$ as $y \sim H^{-1}$ for given values of
the constant $k-$essence potential $V$.
The temporal behaviour of quantities like $a, \dot{a}, H$ during the
late time cosmic evolution has been extracted from the SNe Ia observations.
The dynamical variables $x$ and $y$ defined  by Eq.\ (\ref{eq:defxy})
capture this observed cosmological dynamics, in the context when the 
dark energy is realised by a  scalar field $\phi$ dynamically driven
by a $k-$essence Lagrangian with constant potential $V$ leading
to the scaling relation $XF_X^2 = Ca^{-6}$ involving a constant $C$.\\

Also note from Eq.\ (\ref{eq:aa15}) that, in the
expression for $\sqrt{X}$, ($1/\alpha$) 
appears as a constant multiplicative  factor with the temporal
part known from SNe IA observation. Since, $x=\dot{\phi}=\sqrt{2X}$, the constant 
$\alpha$ may be absorbed in the definition of the field $\phi$ (and hence in $x$) 
by the rescaling
$\phi \to \phi/\alpha$. However, as can be seen
from Eq.\ (\ref{eq:aa17}), the effect of the constants $C$ and $V$ explicitly
show up in the functional form of $F(X)$ extracted from the observational 
data by virtue of the scaling relation. This feature is reflected in the
difference in the phase orbit plots (presented in figs.\ \ref{fig:ppap1}, \ref{fig:ppa1} and \ref{fig:ppa10})
of the representative autonomous system  for different benchmark choices of
values of $(C, \alpha)$ or $(C,V)$. \\

Finally we describe the significance of the four stable points 
occurring for a given choice of the values of  $(C,\alpha)$,
in the context of dark energy scenario represented by a scalar field $\phi$ with a constant potential $V$
in its Lagrangian.
As mentioned earlier the value of equation of state of the total dark 
fluid $w$ is -1 at all the stable points implying approach of
the phase orbits towards all the stable attractors correspond  to acceleration
of expanding universe. The four stable points lie in the 4 different quadrants in the
$x-y$ plane. For both the points in 1st and 4th quadrant we have $x = \dot{\phi}  > 0$.
This implies that,   the orbits approaching towards these two stable points
correspond to the scenario where the value of the real scalar $k-$essence field 
increases monotonically with time as the current universe is accelerating. On the other hand,
for both the points in 2nd and 3rd quadrant we have $x  =\dot{\phi}  < 0$, implying 
realisation of approaching of orbits towards these two stable points by
a   scenario where the value of the $k-$essence field 
decreases monotonically with time. So both scenarios of the homogeneous
scalar field $\phi(t)$ increasing with time and decreasing with time
can be accommodated in JLA data. However, each of the two stable points for
each of above-discussed scenarios correspond to $y>0$ (for the point in upper half 
plane) and $y<0$ (for the lower half plane and also, as discussed earlier,
these points are symmetrically positioned about the $x-$axis. The reason 
for these, as seen from the definition of the dynamical variable $y = \kappa \sqrt{V} / \sqrt{3} H$,
is that for a given choice of $V$, both the positive square root $(\sqrt{V})$ and
the negative square root $(-\sqrt{V})$ affects the description of the representative 
autonomous system through the dynamical variable $y$. But the dark energy model and its
cosmological consequences remain insensitive to the sign of $\sqrt{V}$, as only $V$ 
(and not $\sqrt{V}$) appears in the   phenomenological form
of $F(X)$ extracted from observation which enters in the equation of motion 
of the $k-$essence field in Eq.\ (\ref{eq:aa12}). The occurrence of the
two stable points for a given $x = \dot{\phi}$, one in 
upper half plane (with $\sqrt{V}>0$) and the other in the lower 
half plane (with $\sqrt{V}>0$) is thus only an artefact arising from 
the choice of the specific
autonomous system, for the purpose of realising the interplay of
the constants $C$ and $V$ in the context of JLA data and the dark energy model
considered here.

\section{Conclusion}
\label{sec:con}
In this paper, we considered a homogeneous $k$-essence scalar field $\phi$
with a non-canonical Lagrangian of the form 
$L=VF(X), X = (1/2)\dot{\phi}^2$ 
with constant potential $V$ governing the  
dynamics of the dark energy of the universe during its late time
phase of evolution.
The constancy
of the potential ensures existence of a scaling relation 
$XF_X^2 = Ca^{-6}$ ($C$ = constant) which connects
the dynamical term $F(X)$ of the Lagrangian with the 
scale factor $a(t)$ of the expanding universe, with a flat 
FRW spacetime metric filled with perfect fluids. Using 
the Supernova Ia observations (JLA data), which is instrumental
in probing features of  late time cosmic evolution, we obtain
the temporal behaviour of the cosmological quantities like 
the scale factor $a(t)$, the equation of state and energy density 
of the dark fluid content (dark matter dust and dark energy)
of the universe. Using these results 
obtained from the model independent analysis of the JLA data,
we obtain the time dependence of the $k$-essence
scalar field and reconstructed the $X$-dependence
of the dynamical term $F(X)$ in the Lagrangian.
In our analysis,  we ignore the contribution
from radiation and baryonic matter to the energy density of
the universe during its late time phase of evolution
based on the observations of  present-day relative
densities of different components from Planck collaboration.  
The value of the constant potential $V$ and the constant $C$
appearing in the scaling relation are parameters of this
model.\\

To investigate the interplay of the values of the parameters
$C$ and $V$ in controlling the cosmological dynamics in
the context of the constant potential $k-$essence model and the JLA data,
we use the mapping of the model to dynamical system represented
by a set of
autonomous equations involving two dimensionless variables
$x$ and $y$ suitably defined in terms of cosmological
quantities and relevant parameters of the model.
We investigate impact
of the parameter values $(C,V)$ on the analysis of this dynamical
system. For any specific choice of values of parameter set $(C,V)$,
the system  contains 
the observational inputs from JLA data.
For convenience of calculations involved, we choose
an equivalent parameter set $(C,\alpha)$ where $\alpha \equiv
2\sqrt{C}V/(\rho_{\rm de}^0 + \rho_{\rm dm}^0)$, to present the
results of our analysis. We used linear stability theory
to analyse dynamical features of the system and 
for any ($C,\alpha$) we obtain 
six real fixed
points of the system two of which are saddle points 
and the other four being stable fixed points. We find
that the two saddle fixed points correspond to a  fixed point
solution $y=0$ and lie on the $x-$axis.
At the saddle points, the equation of 
state of the total dark fluid $w=0$  and they do not
correspond to the accelerating universe.  
The four stable fixed points correspond
to   accelerating universe with $w=-1$ and
they may be grouped into two pairs, where points of each pair 
have $y-$values of same magnitude with opposite signs
as evident from the fixed point solution given in
Eq.\ (\ref{eq:ysol2}) which has a $y \to -y$ symmetry.
We investigated how locations of all these fixed points
in $x-y$ plane change with change in the values of  
parameters: $C$ and $\alpha$. In the context of  
$k$-essence cosmological model of dark energy and the observational JLA data,
this provides an indirect 
approach towards realising the dependence of  
dynamical aspects of the model 
on the values of the constant potential $V$ and the constant $C$
in the scaling relation.\\

Note that, the approach is taken here in considering the SNe Ia data and associated errors are simplistic and restricted to a particular class of modes which does not allow uncertainties at smaller redshifts as can be seen from Fig.\ \ref{fig:1}. We avoided a rigorous and comprehensive  
analysis of the SNe Ia data and took this simplistic approach obtaining a gross profile of 
some cosmological parameters from observation, as this provides the optimal, necessary 
observational inputs required to emphasize what we explored in this work. \\

Our primary focus, in this paper, is to investigate the role of values of two constants 
\textit{viz.} $C$ and $V$ in controlling dynamical features of late time dark energy dominated 
era cosmic evolution,  where the constant $C$ appears in the scaling relation arising in the 
context of $k$-essence model of dark energy driven by a Lagrangian with a constant potential $V$.  
In this paper we wanted to show the results of such an investigation  by mapping dynamical 
aspects of the k-essence model onto a two-dimensional dynamical system driven by a set of 
autonomous equations involving the model parameters $V$ and $C$. The variation of corresponding 
resulting stable points and the phase orbits with values of $V$ and $C$ has been shown - which reflects the sensitivity of the dynamical features of the $k$-essence model on $V$ and $C$. 
While taking the different benchmark values of $V$ and $C$ to depict such variations, 
we ensured that their chosen values  correspond to late time profile of relevant 
cosmological parameters which are more or less compatible with their extracted profile from JLA data. \\

In this paper, we wanted to show the results of such an investigation  by mapping the dynamical aspects 
of the k-essence model onto a two-dimensional dynamical system driven by a set of autonomous 
equations involving the model parameters $V$ and $C$. The variation of corresponding resulting 
stable points and the phase orbits with values of $V$ and $C$ has been shown - which reflects 
the sensitivity of the dynamical features of the $k$-essence model on $V$ and $C$. While taking 
the different benchmark values of $V$ and $C$ to depict such variations, we ensured that their 
chosen values  correspond to late time profile of relevant cosmological parameters which are  
compatible with their  profile extracted from JLA data.  The role of observational input 
is only up to this extent, in the context of the  present work which  emphasizes on the study 
of the sensitivity of the dynamical features of the $k$-essence model on above mentioned parameters 
$V$ and $C$. So we only took with the best-fit values of the different coefficients occurring 
in different parametrisations used in the paper without considering propagation of 
uncertainties to the coefficients in equations and evaluating corresponding covariance matrix.

\paragraph{Acknowledgement}\
We would like to thank the honourable referee for valuable suggestions. A.C. would like to thank Indian Institute of Technology, Kanpur for supporting this work by means of Institute Post-Doctoral Fellowship \textbf{(Ref.No.DF/PDF197/2020-IITK/970)}.

\end{document}